\documentstyle[twoside,fleqn,espcrc2]{article}
\input{epsf.sty}
\def\b{\beta}

\def\d{\delta}
\def\g{\gamma}

\def\k{\kappa}     % Also, \varkappa (see below)
\def\l{\lambda}
\def\m{\mu}

\def\z{\zeta}

\def\cD{{\cal D}}

\def\gp{\gamma^{\prime}}

\newcommand{\av}[1]{\mbox{$\langle #1 \rangle$}}

\newcommand{\half}{\mbox{{\normalsize $\frac{1}{2}$}} }
\newcommand{\quart}{\mbox{{\small $\frac{1}{4}$}} }
\newcommand{\fiha}{\mbox{{\small $\frac{5}{2}$}} }

\newcommand{\ra}{\rightarrow}

\newcommand{\be}{\begin{equation}}
\newcommand{\ee}{\end{equation}}
\newcommand{\bea}{\begin{eqnarray}}
\newcommand{\eeaa}{\end{eqnarray}}
\newcommand{\eq}{\ref}
\newcommand{\beq}{\begin{equation}}
\newcommand{\eeq}{\end{equation}}
\newcommand{\cc}{\cite}
\newcommand{\lb}{\label}
\def \3{\ss}

\newcommand{\AmS}{{\protect\the\textfont2
  A\kern-.1667em\lower.5ex\hbox{M}\kern-.125emS}}

\title{Regge's space-time skeletons and the quantization of 2d gravity
            }

\author{Wolfgang Bock\address{Dept. of Physics,
        University of California, San Diego,
        Gilman Dr.~0319, La Jolla, CA 92093-0319, USA}%
        }
\begin{document}

\begin{abstract}
Regge's method for regularizing euclidean quantum gravity is applied
to two dimensional gravity.  Using topologies with
genus zero and two and a scale invariant measure, we show
that the Regge method fails to reproduce the values  of the   string
susceptibilities of the continuum model.
\end{abstract}

\maketitle
\section{INTRODUCTION}
One possibility to quantize gravity is to focus on finding
a regularization of the euclidean path integral. Most studies
so far have used either the dynamical triangulation (DT) or the Regge
approach (RA). The continuum path integral is replaced in both cases
by a summation over simplicial manifolds (skeletons). In the RA
the edge lengths are the dynamical
degrees of freedom and the connectivity of the skeleton is kept fixed,
whereas vice versa
in the DT the edge length is kept fixed and the summation in the path
integral is over skeletons with different connectivity.
An ideal test ground for the two regularization schemes is
two dimensional gravity where many exact results have been derived
in the continuum.
It has been shown analytically and numerically that the DT
reproduces indeed the results of the continuum theory. Based on
a numerical simulation it has been
claimed in ref.~\cc{GrHa91} that this is also the case for the RA.
It will be shown in the following that this statement is incorrect.
Further evidence for the failure of the Regge approach has recently been
given in ref.~\cc{HoJa}. Let's consider now the  continuum path
integral for 2d pure gravity
\bea
Z(A) \!\!\!\!& = &\!\!\!\!\! \int \frac{\cD g}{{\rm Vol(Diff)}} \; e^{-S(g)} \;
       \d(\int d^2x \sqrt{g} - A) \;, \lb{ZZPART} \\
 S(g)\!\!\!\!& = &\!\!\!\!\!\int d^2x \; \sqrt{g} \; (\l + \k R + \b  \quart
R^2)
 \;. \label{SCONT}
\eeaa
The $\d$ function imposes the constraint that the total area of the
surface is equal to $A$. We shall omit the Newton's term in the
following since it is related to a topological
invariant. It is an irrelevant constant as long as the topology is kept
fixed. On a formal level, i.e.  when
ignoring the measure in (\eq{ZZPART}) one would expect that
$Z(A)\propto \exp[-\l A] /A$.
Quantum fluctuations are seen to lead to deviations
from this naive scaling relation.
Using techniques of conformal field theory Kawai and Nakayama
showed that the regularized path integral obeys for $\b/A \ll 1$
the scaling relation \cc{KaNa}
% FIGURE 1
%
\begin{figure}[tb]
 \centerline{
 \epsfysize=6.5cm
 \epsfbox{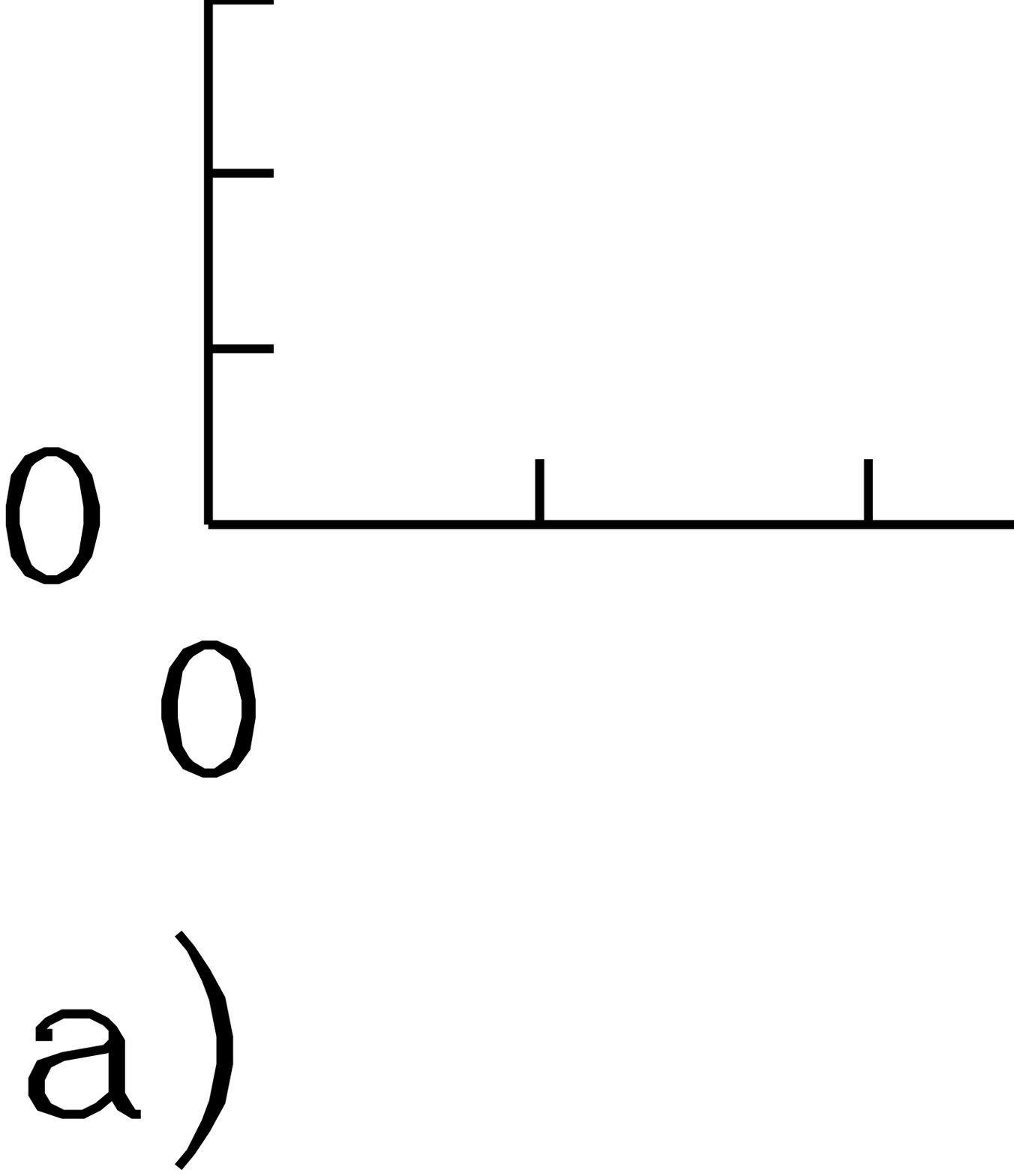}
 }
%\vspace*{0.5cm}
 \centerline{
 \epsfysize=6.5cm
 \epsfbox{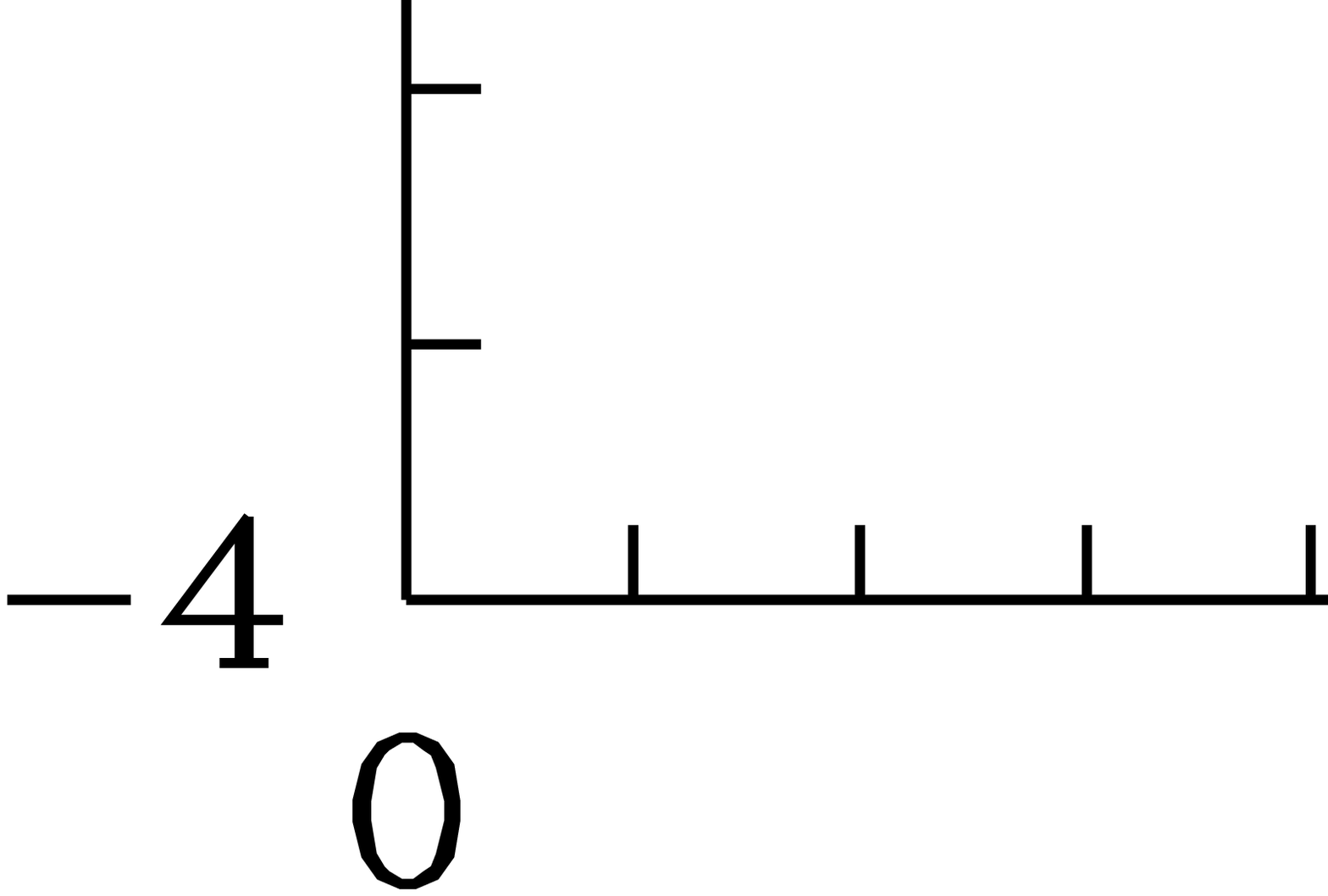}
 }
%\vspace*{5.0cm}
\vspace*{-1.2cm}
\caption{ \noindent {\em $c_1$ for the bi-torus and the sphere.
}}
\label{FIG1}
\end{figure}
\be
 Z(A)\propto A^{\g_{{\rm str}}-2} \exp [   -\l_R A]/A \;,  \label{ZCONTA}
\ee
where $\l_R$ is the renormalized cosmological constant.
The string susceptibility
$\g_{{\rm str}}=2-\fiha (1-h)$ depends on the genus $h$ of the surface.
For $\b/A \gg 1$ they derived the relation
\be
 Z(A)\propto A^{\gp_{{\rm str}}-2} \exp [-S_{R^2}^{{\rm cl}}(A)]
\exp [   -\l_R A]/A \;,  \label{ZCONTB}
\ee
where $\gp_{{\rm str}}=2-2(1-h)$ is a string susceptibility that
is different from $\g_{{\rm str}}$ in
eq.~(\ref{ZCONTA}) and $S_{R^2}^{{\rm cl}}(A)= 16 \pi^2 (1-h)^2 \b/A$ is
the classical action of the $R^2$ part in (\eq{SCONT}).
The renormalization of
the cosmological constant and the emergence of the factors
$A^{\g_{{\rm str}}-2}$ and $A^{\gp_{{\rm str}}-2}$ are due to the
quantum fluctuations.  We will investigate in the following sections
the question if the RA can reproduce the
continuum results for $\g_{{\rm str}}$ and $\gp_{{\rm str}}$.
Since relations (\eq{ZCONTA}) and (\eq{ZCONTB})
coincide for a torus
with the naive scaling relation (except for $\l \ra \l_R$)
we shall consider in the following
only topologies with $h=0$ (sphere) and $h=2$ (bi-torus).
\section{THE REEGE METHOD}
Using the RA we can regularize the path integral in (\eq{ZZPART})
as follows
\bea
Z(A,N_1) \!\!\!\!\! &=& \!\!\!\!\! \int_0^{\infty}\!\!\!\!\!\!
d\m(l)\;F(l)\;e^{-S(l) }
                 \d(\sum_i A_i-A) \;, \lb{RZZ} \\
S(l) \!\!\!\! &=& \!\!\!\! \sum_i(\l \; A_i + \b \; \d^2_i/A_i)
                  \;, \label{ZREGCON}
\eeaa
where the summations are over all vertices $i$ in the skeleton  and
$A_i(l)$ and $\d_i(l)$ are the area and
deficit angle associated with the vertex $i$.
The argument $N_1$ denotes the total
number of links in the skeleton. The factor $F(l)$ in (\eq{RZZ})
ensures that the triangle inequalities are fulfilled.
The integration
is over edge lengths in the skeleton.
A serious problem is that it is not clear which
measure corresponds to the gauge invariant measure in eq.~(\eq{ZZPART}).
A measure that has been chosen so far in literature is of the very
simple form $d\mu (l)=\prod_{k} (d l_{k}^2 /l_{k}^2) l_{k}^{\z}$
which is scale invariant if the parameter $\z$ is equal to
zero.
To study the scaling behavior of the path integral we follow \cc{GrHa91}
and consider the derivative
\be
\frac{d \log Z(A,N_1)}{dA}= -\l + \frac{1}{A} \left( [
S_{R^2} + \frac{N_1\z}{2}] -1 \right),
                \lb{DZDA}
\ee
where $S_{R^2}=\beta \langle \sum_i \d^2_i/A_i \rangle$ and
$\l \b$ and $N_1$ are kept fixed. A derivation of
relation (\eq{DZDA}) has been  given in ref.~\cc{BoVi}.
The scaling formulas (\eq{ZCONTA}) and (\eq{ZCONTB}) can
only be reproduced if the
sum of two terms in square brackets turns in the infinite volume limit,
$N_1 \ra \infty$, into expressions of the form $-(\l_R-\l) A +
\g_{{\rm str}}-2$ and $-(\l_R-\l) A +
\gp_{{\rm str}}-2 + S_{R^2}^{{\rm cl}}(\b/A)$.
The average action $S_{R^2}$ is
an extensive quantity and therefore
both terms in square brackets in (\eq{DZDA}) are $\propto N_1$.
This implies that $\z$ should be chosen
such that it cancels the term in $S_{R^2}$ which is
$\propto N_1$. Already at this stage strong doubts arise
that the RA can reproduce the continuum results for $\g_{{\rm str}}$
and $\gp_{{\rm str}}$.
Instead of the measure $d \mu (l)$ we could namely have used
also a different measure where the exponent $\z$ is replaced  by $\z_0 +
\z_1/N_1$. An expansion to leading order in $\z_1/N_1$ shows
that $\g_{{\rm str}}$ and $\gp_{{\rm str}}$ get shifted by
$\z_1(\half+C)$ with $C$ the connected part of
$\av{[\sum_{k}\log l_{k}][\sum_j A_j (\b/A) \sum_i \d_i^2/A_i]}$,
evaluated for $\z_1=0$.
It is very unlikely that $C$ is exactly equal to $-1/2$, such that
the string susceptibilities remain unchanged. Moreover for $\z_0=\b=0$
one finds $\g_{{\rm str}} = \gp_{{\rm str}} = 2+\z_1/2$ which is
independent of the genus $h$ and completely arbitrary.
\section{NUMERICAL RESULTS}
These considerations cast serious doubts on the claim made in ref.
\cc{GrHa91}
that the Regge model with scale invariant measure reproduces the
continuum value of the string susceptibility $\g_{{\rm str}}$.
To settle this, we
have computed $S_{R^2}$ numerically for
the sphere ($h=0$) and the bi-torus ($h=2$)
using the scale invariant measure, i.e. for $\z=0$.
We have constructed the sphere from the
surface of a three dimensional cube. The bi-torus has been obtained
by gluing together two tori along the
boundary of a cut out window. Details are given in
ref.~\cc{BoVi}. Two different algorithms have been used to simulate the
fixed area path integral (\eq{RZZ}): a hybrid Monte Carlo algorithm
and a Monte Carlo algorithm combined  with
a histogramming method. The data generated with the two  methods
agree within error bars.
%
%
% FIGURE 2
%
\begin{figure}[tb]
 \centerline{
 \epsfysize=6.5cm
 \epsfbox{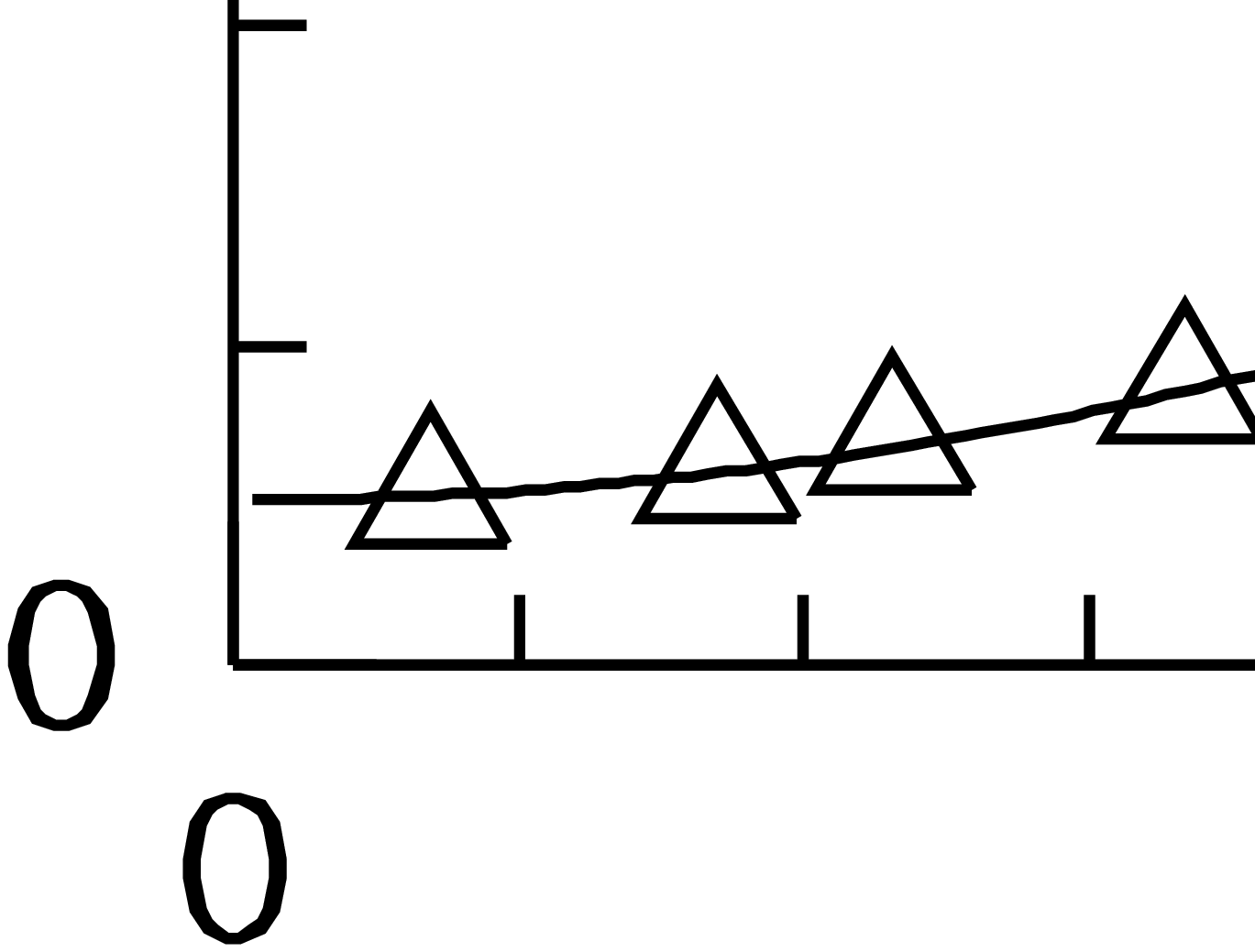}
 }
%\vspace*{5.0cm}
\vspace*{-1.2cm}
\caption{ \noindent {\em $S_{R^2}$ as a function of $1/N_2$.
}}
\label{FIG2}
\end{figure}
\\
\noindent \underline{$\b/A \ll 1$}: In order to determine the string
susceptibility $\g_{{\rm str}}$ we have to
compute the average action $S_{R^2}$
as a function of the total area $A=a N_2$, where
$N_2$ is the number of triangles in the skeleton.
Because of technical resons we have kept $\b/a$ fixed and
varied $N_2$. The case $\b/A \ll 1$ is then
realized for large values of $N_2$ and small values of $\b$.
The numerical results for $S_{R^2}$ could be fitted in all cases to an
ansatz of the form $S_{R^2}=c_0(b/a)N_2+c_1(\b/a)$.
It has been shown in ref.~\cc{BoVi} that $\g_{{\rm str}}-2$ is then
given by $\lim_{\b/a \ra \infty} c_1(\b/a)$. Fig.~1a (bi-tours)
and b (sphere) show that for the sphere and
the bi-torus the numerical estimates for $\g_{{\rm str}}-2$
are substantially larger than the continuum results which are
represented by the horizontal dashed lines.
The triangle in fig. 1a has been obtained after adjusting $\z$
such that the term $\propto N_1$ in (\eq{DZDA}) cancels.
It occurs that the point falls
on top of a fit curve through the data points
obtained for $\z=0$ (see ref.~\cc{BoVi}).
This result indicates that $c_1(\b/a)$ does not depend on $\z$. \\
\noindent \underline{$\b/A \gg 1$}: This case is realized
for large values of $\b$ and small values of $N_2$.
We find that the average action can be fitted to an ansatz of the
form $S_{R^2}=c_0(b/a) N_2 +c_1(\b/a)+c_2(\b/a)/N_2$.
An example for such a fit
is shown in fig.~2 for $\b/a=100$. The string susceptibility
$\gp_{{\rm str}}$ and the quantity
$S_{R^2}^{{\rm cl}}/(\b/A)=16 \pi^2 (1-h)^2$
are now given by the limites
$\lim_{\b/a \ra \infty} c_1(\b/a)$ and
$\lim_{\b/a \ra \infty} c_2(\b/a)/(\b/a)$. It turns out that
$c_2$ can be determined with higher accuracy than
$c_1$. This is understandable because the first  exponential
in (\eq{ZCONTB}) dominates the scaling behavior.
Some results for
$c_1(\b/a)$ and $c_2(\b/a)/(\b/a)$ are given in the following
table:
\vspace*{-1.0cm} \\
\begin{table}[hbt]
% space before first and after last column: 1.5pc
% space between columns: 3.0pc (twice the above)
\setlength{\tabcolsep}{.41pc}
% -----------------------------------------------------
% adapted from TeX book, p. 241
\newlength{\digitwidth} \settowidth{\digitwidth}{\rm 0}
\catcode`?=\active \def?{\kern\digitwidth}
% -----------------------------------------------------
%\caption{$c_1$ and $c_2/(\b/a)$ as a function of $\b/a$}
\label{tab:lb}
%\begin{tabular*}{\textwidth}{@{}l@{\extracolsep{\fill}}rrrr}
% \begin{tabular*}{\textwidth}{@{}l@rrrr}
\begin{tabular}{|c|c|c|c|c|}
\hline
                 & \multicolumn{2}{c|}{sphere}
                 & \multicolumn{2}{c|}{bi-torus} \\ \cline{2-5}
$\b/a$           & $c_1$ & $c_2/(\b/a)$ & $c_1$ & $c_2/(\b/a)$    \\ \hline
$50$             & $-1.9(5)$ & $158.9(4)$ & $-0.7(2)$ & $157.4(2)$ \\
$70$             & $-1.9(5)$ & $158.6(4)$ & $-0.7(3)$ & $157.8(2)$ \\
$100$            & $-2.2(8)$ & $158.6(4)$ & $-0.8(4)$ & $157.6(2)$
\\
\hline
% \multicolumn{5}{@{}p{120mm}}{Reprinted from: G.M. Ritcey,
%                              Tailings Management,
%                              Elsevier, Amsterdam, 1989, p. 635.}
\end{tabular}
\end{table}
\vspace*{-0.8cm} \\
The data for $c_1$ and $c_2/(\b/a)$
seem not to depend very much on $\b/a$
and are presumably  close their $\b/a
=\infty$ extrapolations.
The results for $c_2/(\b/a)$ are for both topologies
close to $16 \pi^2=157.91$...
Also
the coefficient $c_1$ agrees for the sphere
within the large error bars with the continuum result
$\gp_{{\rm str}}-2=-2$. In the case of the bi-torus the deviation
from the continuum value
$\gp_{{\rm str}}-2=2$ is however again substantial. \\
The results reported in the last two sections strongly indicate
that the RA fails to  reproduce the quantum effects in 2d
$R^2$ gravity. Only the classical term in (\eq{ZCONTB}) and
perhaps $\gp_{{\rm str}}$ for the sphere are reproduced
correctly.
A likely explanation for the failure of the RA is
the contribution of gauge degrees of freedom which appear
not to decouple from the path integral.

Most of the results have been obtained in a collaboration
with J.C. Vink. I would like to thank J. Kuti, J. Smit and J.C. Vink
for many interesting discussions.


\begin{thebibliography}{9}
\bibitem{GrHa91} M. Gross, H.W. Hamber, Nucl. Phys. B364 (1991) 703.
\bibitem{HoJa}   C. Holm,    W. Janke, Phys. Lett. B335 (1994) 143.
\bibitem{KaNa}   H. Kawai, R. Nakayama, Phys. Lett. B306 (1993) 224.
\bibitem{BoVi}   W. Bock,    J.C. Vink, UCSD/PTH 94-08.
\end{thebibliography}
\end{document}